%% file: paper.tex
\documentclass[runningheads]{llncs}

\usepackage{graphicx}
\usepackage{listings}
\usepackage{xcolor}
\usepackage{xspace}

\newcommand{\consult}{\textsc{Consult}\xspace}

\usepackage{glossaries}
\loadglsentries{glossary}

\begin{document}
\title{Non-repudiable provenance\\
for clinical decision support systems
\thanks{This work has been supported by European Union’s Horizon 2020 research and innovation programme under grant agreement No 654248, project CORBEL, and under grant agreement No 824087, project EOSC-Life.}
}
%
%

\author{Elliot Fairweather\inst{1}
\and
Rudolf Wittner\inst{2,3}
\and
Martin Chapman\inst{1}
\and
Petr Holub\inst{2,3}
\and
Vasa Curcin\inst{1}
}

\authorrunning{Fairweather, E. et al.}

\institute{Department of Population Health Sciences\\ King's College London, United Kingdom\\
\email{elliot.fariweather@kcl.ac.uk}\\
\and
BBMRI-ERIC, Graz, Austria
\and
Institute of Computer Science \& Faculty of Informatics \\
Masaryk University, Brno, Czech Republic\\
\email{rudolf.wittner@bbmri-eric.eu}
}

\maketitle

\begin{abstract}
Provenance templates are now a recognised methodology for the construction of    
data provenance records.  Each template defines the provenance of a              
domain-specific action in abstract form, which may then be instantiated          
as required by a single call to the provenance template service.
As data reliability and trustworthiness becomes a critical issue in an increasing
number of domains, there is a corresponding need to ensure that the provenance of
that data is non-repudiable.         
In this paper we contribute two new, complementary modules to our template model 
and implementation to produce non-repudiable data provenance.  The first, a      
module that traces the operation of the provenance template service itself, and            
records a provenance trace of the construction of an object-level document,      
at the level of individual service calls. The second, a non-repudiation          
module that generates evidence for the data recorded about        
each call, annotates the service trace accordingly, and submits a representation of that evidence to      
a provider-agnostic notary service.                                              
We evaluate the applicability of our approach in the context of a clinical       
decision support system.
We first define a policy to ensure the non-repudiation of evidence
with respect to a security threat analysis
in order to demonstrate the suitability of our solution.
We then select three use cases from within a particular system, \consult, with contrasting data provenance          
recording requirements and analyse the subsequent performance of our prototype   
implementation against three different notary providers.   
\end{abstract}

\section{Introduction}

The \textit{provenance} of a data resource describes the entities, activities and agents that have influenced it over time~\cite{moreau2013prov}. \textit{Provenance templates} are a methodology for the construction of data provenance records. A template is an abstract provenance document which may be later instantiated, as many times as required, usually in the context of a larger document, to produce a concrete provenance document~\cite{Curcin2017}. Each template is designed to represent a discrete, domain-specific action which can be recorded as a single call to a provenance template service~\cite{templates-service}.
Provenance tools are now being used widely in scientific domains, where \textit{trust} in the provenance records constructed is essential~\cite{Moreau2017}. A lack of transparency, which implies a lack of trust, is considered one of the main reasons for the poor uptake of clinical decision support systems (\gls{decision support system}) \cite{10.1197/jamia.M3170}. In response to this issue, there has been a recent movement towards \textit{secure} provenance, for example \cite{Massi2018}.

One key security objective is \textit{non-repudiation}, which is defined as \textit{preventing the denial of previous commitments or actions}~\cite{Menezes:1996:HAC:548089}. In this paper, we focus specifically on non-repudiation of origin -- preventing an author from falsely denying the act of creating content or sending a message -- referred to simply as non-repudiation in the remainder of the text. 

Whilst the correctness of the operations commonly used to achieve non-repudiation at the cryptographic level can be formally proven, at the practical level, the broader context needs to be taken into account,
such as international regulations related to digital signatures and communications, or current threats relevant to specific systems, which evolve over time.
For example, applying a digital signature to a message, which is a common way to ensure its authenticity, means only that a cryptographic operation was performed by a piece of hardware or software on behalf of a person. In a real-world scenario, we might then presume that that person is the real author of a message, because we have assumed that the corresponding private key is known only to them. However, this might not hold true; there is always a chance that confidentiality of a private key can be violated (for example, by malware) and there is no guaranteed mechanism to prevent such an event.

To address this challenge, we first perform a threat analysis for the non-repudiation of recommendations made by a \gls{decision support system}, and define a non-repudiation policy which can be later used during an adjudication process. We then use this policy to guide the design of a provenance-based model for the representation of the evidence required for non-repudiation, and implement this as an optional feature within our provenance template service. We then evaluate our solution using three different evidence storage solutions that satisfy our policy requirements, in the context of a \gls{decision support system}, \consult.

\section{Related Work}
\label{sec:related-work}

Authenticity, commonly defined as corroboration of a claimant's identity, has long been considered distinct from non-repudiation~\cite{10007981481,Menezes:1996:HAC:548089}. The former is considered a simpler security requirement than the latter, which is typically a more complex, protocol-based security service~\cite{Hafner:jucs_15_15:seaas_a_reference_architecture} and is also defined as one of nine security principles in \cite{10.1007/3-540-58618-0_67}. The current ISO standard for non-repudiation~\cite{ISO/TCJTC1SC272009} explicitly states that \textit{non-repudiation can only be provided within the context of a clearly defined security policy for a particular application and legal environment}. Detailed discussion about why non-repudiation and related evidence management must be designed in advance can be found in the thesis of Roe~\cite{UCAM-CL-TR-780}.

To our knowledge, secure provenance and related challenges were discussed for the first time in 2007 \cite{Hasan:2007:ISP:1314313.1314318} and were focused on its integrity, availability and confidentiality. Another paper \cite{Braun:2008:SP:1496671.1496675} discussed related challenges in a more detailed way and explained that existing security models do not fit to graph structures, which is the standardized representation of provenance information. 

The use of a notary service within a system to ensure the integrity and non-repudiation of biomedical knowledge retrieval requests and responses from a database has already been implemented~\cite{KLEINAKI2018288} but differs from our solution by employing a so-called in-line notary. This work does not however include an analysis of possible threats or define tactics for their mitigation, which is crucial in the context of achieving non-repudiation. The authors also mention that to the best of their knowledge there currently exists no other work using blockchain-based technology to manage biomedical evidence integrity and non-repudiation. This is supported by a scoping review~\cite{HASSELGREN2020104040} which references the former as the only paper within the domain of healthcare or health sciences to address the property of non-repudiation. 

Another survey paper~\cite{VIGIL201516} describes and compares  existing methods to ensure integrity, authenticity, non-repudiation and proof of existence in the long-term. The authors make no distinction between the terms \textit{authenticity} and \textit{non-repudiation}, which would have dramatic consequences in a real-world use case. The same is true for the proposed secure provenance schemes in~\cite{7847190} and~\cite{JAMIL201834} that claim that non-repudiation can be ensured by applying digital signatures, but any further discussion about what aspects of non-repudiation are achieved is omitted.

\section{A Non-repudiation Policy for Decision Support}
\label{sec:policy}

A clinical decision support system (\gls{decision support system}) is a software tool that evaluates a set of health data inputs and makes recommendations to support clinical decision making. These recommendations range from treatment suggestions to establishing a diagnosis, and are provided to a patient, among other users. Because the recommendation generation process should be transparent~\cite{Curcin2017}, information about that process (\textit{evidence}) is often provided to a patient together with a recommendation, so they can check, for example, whether valid data about their diagnosis and health condition was used (a form of \textit{Explainable AI} (XAI) \cite{Miller2019}). In other words, a patient can, either themselves or via an authorised entity, check the provenance of a recommendation generation process, given that the evidence provided describes the creation and evolution of a particular recommendation. Evidence is particularly useful if harm is caused to a patient as the result of a particular recommendation, perhaps due to an error in the \gls{decision support system}'s implementation or design, when it may, for example, be required in a legal context. In order for evidence to be used in this way, it must be \textit{irrefutable} at the practical level, meaning that it is sufficiently convincing; it is clear that it really came from the system and it was not later changed by anyone to subvert the adjudication process.

We refer to these concerns as \textit{threats}, and identify them within a \gls{decision support system} by examining the following general requirements for irrefutable evidence~\cite{Menezes:1996:HAC:548089}:

\begin{enumerate}

    \item The authenticity and integrity of the provided evidence need to be established, such that the alleged author of the evidence cannot later deny that authenticity.
    
    \item Responsibilities and rules related to evidence generation, storage and verification must be defined in order to enable all participating parties to behave responsibly, in accordance with these rules. Violating these rules can lead to decreased trust in the system. 
    
    \item Authenticity and integrity verification information must be available during a pre-defined period of time, according to a time period for which non-repudiation should be achieved (achieving non-repudiation for an unlimited period, if feasible, is likely to be expensive). The evidence verifier must be able to verify the evidence.
    
    \item If a dispute related to the origin of evidence occurs, trusted timestamps are required to reconstruct past events.
    
\end{enumerate}

When applying these requirements to a \gls{decision support system} we observe, for example, that while a \gls{decision support system} is generally motivated to maintain evidence in order to demonstrate that the decision generation process is compliant with standards and clinical practice, if harm were to be caused to a patient, a \gls{decision support system} creator or owner then has a motive to falsify, modify or destroy generated incriminating evidence -- such as timestamps, in order to discredit the reconstruction of past events -- to protect themselves, particularly if a corresponding authority, responsible for maintaining said standards and practice, is involved. There are also external entities with a motive to disrupt evidence, such as insurance companies, who may wish to avoid making a payment to an injured patient or a \gls{decision support system} provider.

To address these threats, we define a non-repudiation \textit{policy},
given as a set of mandatory requirements for the evidence generation process,
in order to ensure that the evidence produced by a \gls{decision support system} both engenders trust, and can also be later used by a patient during an adjudication process in case of a future dispute. These requirements are presented along with brief rationale:

\begin{enumerate}

    \item\label{req:generation} A \gls{decision support system} is the only party able to generate valid evidence, thus all of evidence needs to be digitally signed. \textit{Authenticity and integrity of the evidence is achieved by this rule}. 
    
    \item\label{req:trusted-hw} Cryptographic operations performed by the \gls{decision support system} are realised using a special piece of hardware, which provides additional protection against private key disclosure. \textit{The reason for applying this rule is that protection of the key needs to be established.}
    
    \item Despite additional protection of private keys, there is always a chance that the confidentiality of a particular private key is compromised. \textit{For that reason, a private key owner should define a certain amount of time to report a confidentiality violation of the private key used for creating digital signatures (this time period is called the \textit{clearance period} \cite{Menezes:1996:HAC:548089}). If the clearance period expires, all digital signatures created before are considered valid. Since this is a processional part of the non-repudiation and it is not important from architecture point of view, we will not address it in our solution.}
    
    \item\label{req:tsa} A timestamp describing the token generation must be generated by a trusted third party. \textit{Integrity and authenticity of the timestamp need to be ensured. Due to the importance of the timestamp, we propose this measure in order to build additional trust in reconstructed events, especially if a signing key had been revoked or expired before a dispute arose.}
    
    \item\label{req:backup} Because a \gls{decision support system} could have a motive to disrupt or destroy existing evidence, it cannot be the only party responsible for storing and maintaining it.
    
    \item\label{req:tamperevident-storage} The evidence used should be held in storage that is able to prevent the modification of stored content. \textit{The reason for applying this rule is to raise the level of trust concerning evidence integrity}.
    
    \item\label{req:token-header} All information needed for integrity and authenticity of the evidence verification should be sealed as part of the evidence. \textit{This rule is intended to simplify additional evidence management}.
    
    \item\label{req:patient-storage} After a decision is made and particular provenance is generated, the patient can verify its meaningfulness, authenticity and integrity, and can store it for prospective future claims. \textit{The assumption here is that the non-repudiation of the provenance is in the best interest of the patient, since their health condition is affected by particular decision. This reflects a current movement in healthcare whereby patients are custodians of their own data \cite{GORDON2018224} and reduces trust assumptions about a third party, which would otherwise have to verify it instead of the patient.}
    
    \item\label{req:patient-id} The generated evidence must contain an identifier for the particular patient. \textit{By applying this rule, the patient may be certain that a generated decision and its evidence was not intended for a different person.}
    
    \item\label{req:additional-timestamps} The evidence generated during a decision generation contains additional information about time when the request from user and when other inputs for the decision generation were obtained.

\end{enumerate}

This policy is used in the following sections to define a secure process for evidence generation, storage and verification within a \gls{decision support system}.
In the following section, we show how some of these requirements can be addressed as a
provenance-based model.

\section{A Provenance-based Model for Non-repudiable Evidence}

\label{sec:nr-model}

The granularity of the provenance template methodology fits the design of an
architecture for non-repudiable evidence perfectly. The fact that each template represent a single,
yet complete, domain-specific action within the client system allows evidence to be
generated and presented at a meaningful, yet manageable scale.
However the direct use of provenance data generated from templates as evidence is insufficient,
because it provides no information regarding its use within the construction of the parent document.
Thus in order to achieve non-repudiation within a client domain, we have first
added the facility to record a provenance trace of the workflow of the provenance
template service itself; we call this data \textit{meta-provenance}.

The meta-provenance trace generated for a document constitutes a wholly reproducible record
of the construction of that document and is described in detail in Subsection~\ref{subsec:metaprovenance}.
This data, later appended with records of the non-repudiable evidence generated at each point in
the workflow as formalised in Subsection~\ref{subsec:nr-evidence}, also allows us to present 
clients with a comprehensive survey of provenance actions carried out on their
behalf and to later validate their authenticity. 

\subsection{Templates for Meta-provenance}
\label{subsec:metaprovenance}                                                   

Each call to the provenance template service, besides some necessary administrative
functionality, represents an action to be executed in the life-cycle of a provenance
document under construction.         
We begin, therefore, by recalling the typical workflow for the construction of a document
using the service in Figure~\ref{fig:server-workflow}.

\begin{figure}                                                                   
\begin{enumerate}                                                                
\item \textit{newTemplate} (one or more times) to upload templates to the server    
\item                                                                            
\begin{enumerate}                                                                
    \item \textit{newDocument} (once) to begin a new document                    
    \item \textit{addNamespace} (zero or more times) to add a namespace to the document
    \item \textit{registerTemplate} (one or more times) to associate a template with the document
    \item \textit{generate} (one or more times) to use a substitution to instantiate a template,
        and merge it into the document                                           
    \item                                                                        
    \begin{enumerate}                                                            
        \item \textit{generateInitialise} (once) to use a subsitution to instantiate a template with
            zones as a fragment                                                  
        \item \textit{generateZone} (zero or more times) to use a substiution to instantiate a
            subsequent iteration of a zone within the fragment                   
        \item \textit{generateFinalise} (once) to check and merge the fragment into the document
    \end{enumerate}                                                              
\end{enumerate}                                                                  
\end{enumerate}                                                                  
\caption{How to construct a document using the provenance template service}                 
\label{fig:server-workflow}                                                              
\end{figure}                                                                    

We now describe how to formalise instances of this workflow and other important
meta-data relating to the operation of the service, by recording provenance traces of
the execution of the actions contributing to the life-cycle of a document. 
Recording the life-cycle of documents not only produces         
valuable meta-data regarding the construction of a document, but is a necessary  
prerequisite to enabling the non-repudiation and later verification of the data
generated.

\begin{figure}[ht!]                                                              
\begin{minipage}[b]{0.5\linewidth}                                               
    \includegraphics[width=0.85\textwidth]{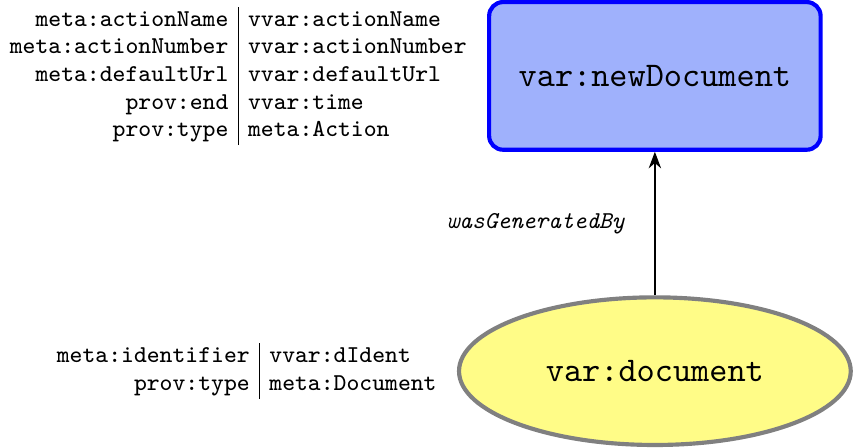}
\end{minipage}                                                                   
\begin{minipage}[b]{0.5\linewidth}                                               
    \includegraphics[width=0.85\textwidth]{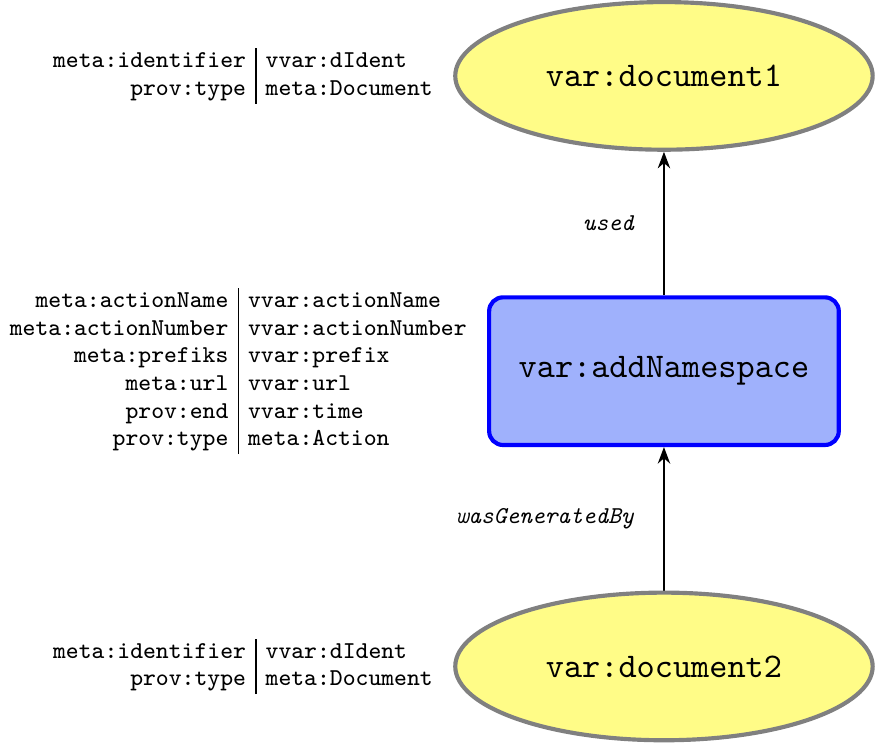}
\end{minipage}                                                                   
\caption{newDocument (\textit{left}) and addNamespace (\textit{right}) templates}                         
\label{fig:template-newdocument}                                                                 
\end{figure}

These traces are recorded in meta-level \textit{history} documents,
one for each standard or \textit{object-level} document under construction.
History documents are valid provenance documents created and maintained by the    
new meta-provenance module, and stored within the server providing the service.                        
Following the creation of new object-level document by the the document          
management module, the meta-provenance module will create a new history document.
Each service call has a corresponding provenance template defined within the     
meta-provenance module.  When the web service receives a request, following a    
successful execution of the request at the object-level by the document          
management module, the request data is sent to the meta-provenance module, which 
builds a substitution for the respective template. This is then instantiated     
within the corresponding separate history document. The templates for the        
\textit{newDocument} and \textit{addNamespace} service calls are shown by way of 
example in Figure~\ref{fig:template-newdocument}.                                                

The meta-provenance templates use annotations in the \texttt{meta} namespace.    
The object document being tracked is represented as an entity of type            
\texttt{meta:Document} and each action executed upon that    
document as an activity of \texttt{meta:Action} type. Each action is annotated   
with its name (\texttt{meta:actionName}) and given a numeric value               
(\texttt{meta:actionNumber}), corresponding to its order in the trace. The time  
that the execution of the action was completed is also recorded as the end time  
of the activity.\footnote{The provenance template model adds three special attributes    
(\texttt{start}, \texttt{end}, \texttt{time}) to the \texttt{prov} namespace in  
order to allow the start and end times of activities, and the times of           
influences to be instantiated as template value variables. These attributes are  
translated in the document model into the respective PROV timings. This is       
necessary because the PROV data model only allows these timings to be of type    
\texttt{xsd:dateTime} and so cannot be replaced by a variable name directly.}    
Templates are recorded as entities of type \texttt{meta:Template} and fragments  
as entities of \texttt{meta:Fragment}. Object documents, templates and fragments 
are all annotated with the \texttt{meta:identifier} attribute which contains     
their unique document identifier within the server.                              
                                                                                 
\begin{figure}[ht!]                                                              
\begin{minipage}[b]{0.5\linewidth}                                               
    \includegraphics[width=0.85\textwidth]{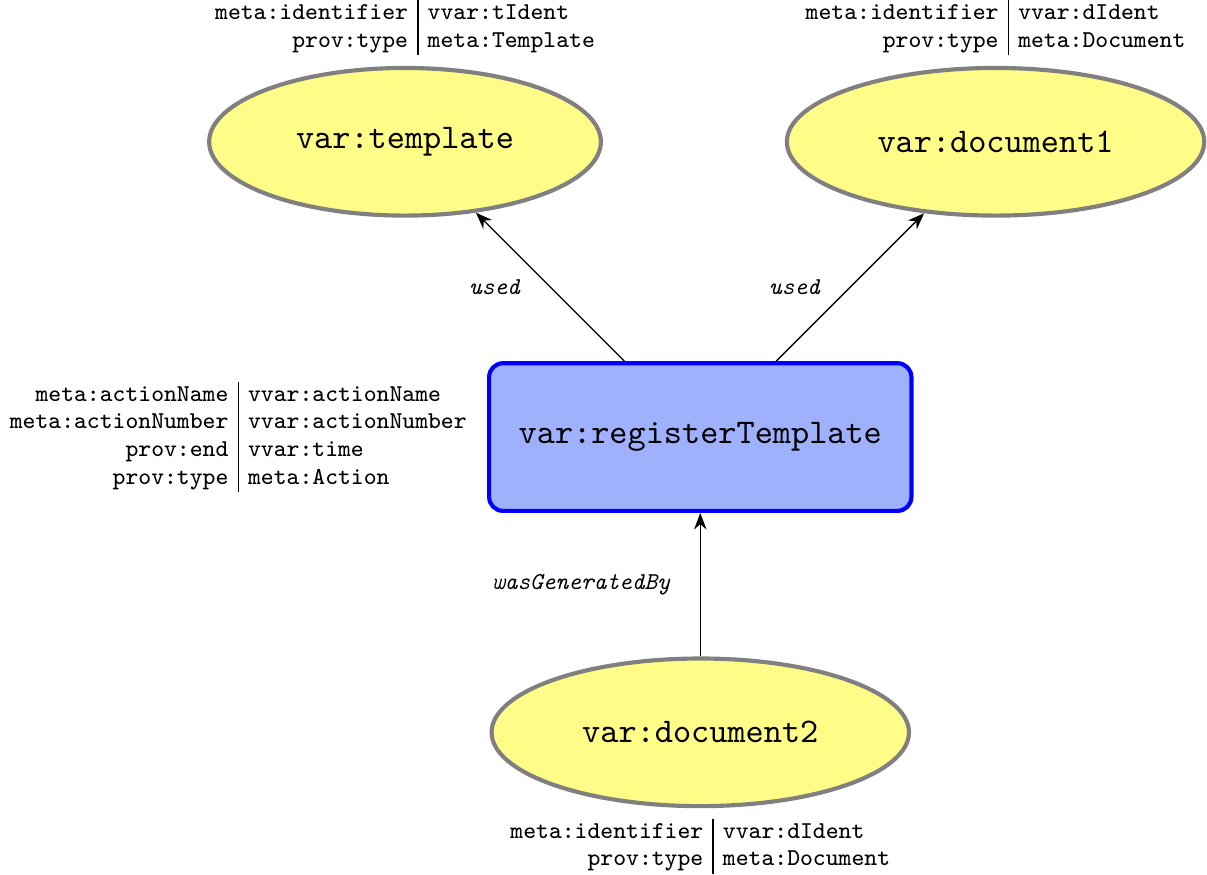}
\end{minipage}                                                                   
\begin{minipage}[b]{0.5\linewidth}                                               
    \includegraphics[width=0.85\textwidth]{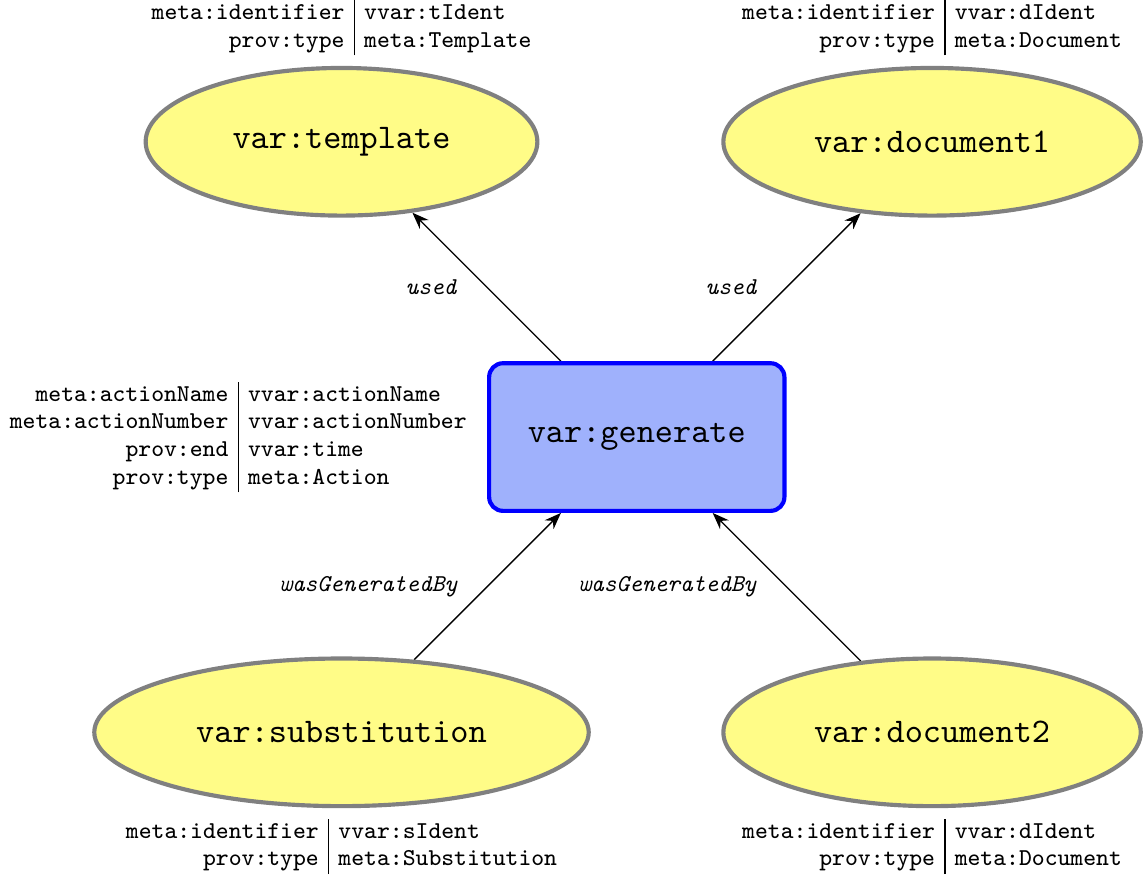}
\end{minipage}                                                                   
\caption{registerTemplate (\textit{left}) and generate (\textit{right}) templates}
\label{fig:template-generation1}                                                                   
\end{figure}                                                                    

\begin{figure}[ht!]                                                              
\begin{minipage}[b]{0.5\linewidth}                                               
    \includegraphics[width=0.85\textwidth]{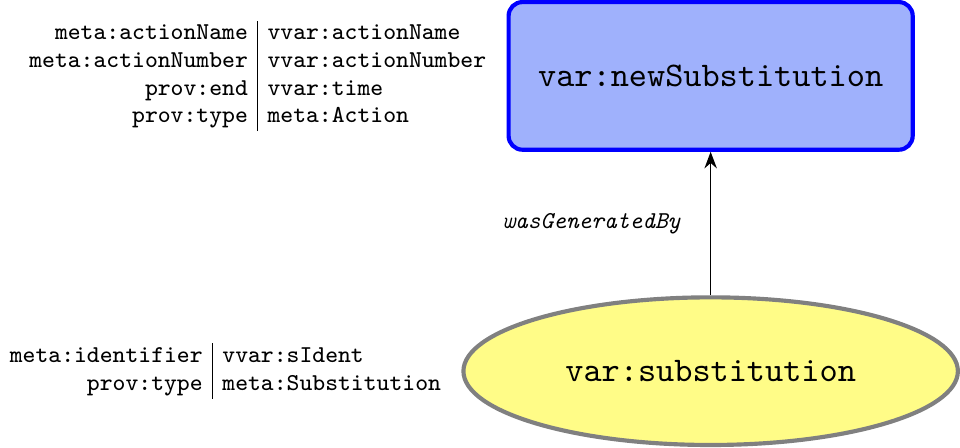}
\end{minipage}                                                                   
\begin{minipage}[b]{0.5\linewidth}                                               
    \includegraphics[width=0.85\textwidth]{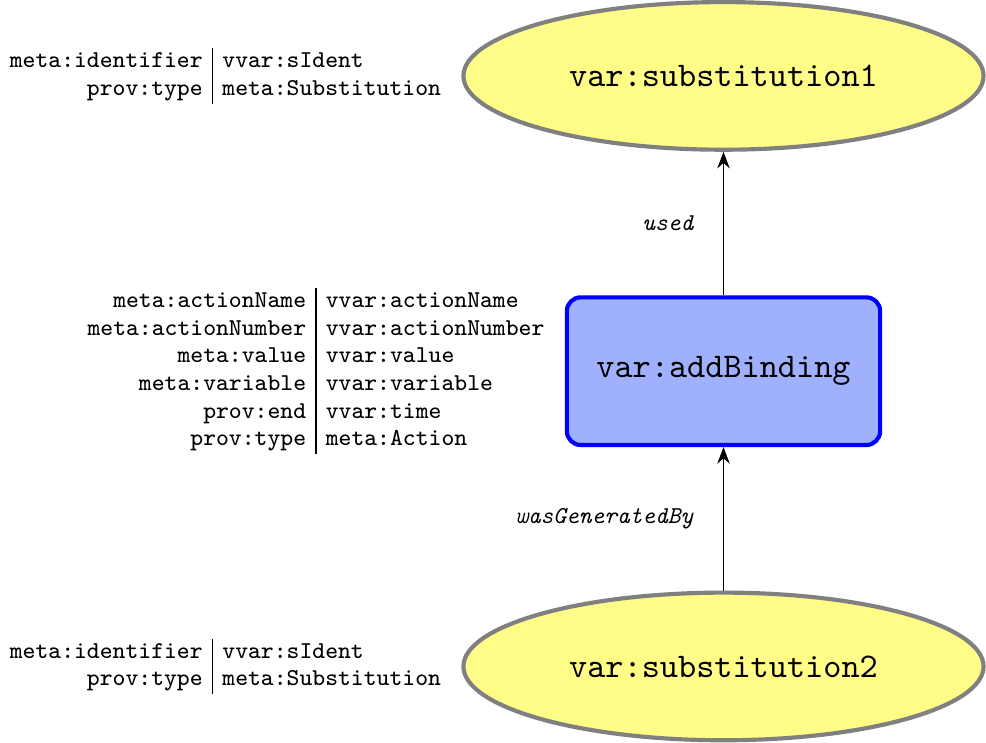}
\end{minipage}                                                                   
\caption{newSubstitution (\textit{left}) and addBinding (\textit{right}) templates}                       
\label{fig:template-substitution}                                                           
\end{figure}

Substitutions given as input to generation actions, that is, \textit{generate},  
(see Figure~\ref{fig:template-generation1})
\textit{generateInit}, or \textit{generateZone}
are now also persisted as meta-level provenance        
documents called \textit{substitution} documents.
A history document together with its associated
substitution documents together form the complete, reproducible meta-procenance record
of the construction of an object document.
The translation of standard   
substitutions into substitution documents is carried out by the meta-provenance  
module again by use of templates, given in Figure~\ref{fig:template-substitution}. A        
substitution is created using the \textit{newSubstitution} template, and then    
each binding added using \textit{addBinding}. These operations instantiations    
are executed in-memory by the server, and the final document is persisted by the 
server, under a system-generated identifier. The substitutions documents thus     
created during the recording of generation actions are referenced in the         
meta-provenance templates as entities of type \texttt{meta:Substitution} and      
again, being valid documents, annotated by their identifier. The fine granularity
of these operations anticipates future improvements to the provenance template service,
whereby substitutions may be submitted to the service over time down to the level
of a single variable binding.
                                                                                 
Any remaining data provided as part of an action,                                
document, is stored as annotations upon the action activity, given in the        
\texttt{meta} namespace. The \texttt{meta:defaultUrl} attribute in the                 
\textit{newDocument} template shown in Figure~\ref{fig:template-newdocument} is one such         
example.                                                                         
                                                                                 
A meta-provenance record is sufficient to reconstruct an object-level provenance  
document in its entirety.
In order to reproduce the construction of a particular document, its history document
is first exported, and the chain of recorded actions then replicated with
reference to the necessary substitution documents.  This facility allows object
documents to be   
recorded by the server at the meta-level alone, to be expanded at a later date.  
After the fact document reconstruction from meta-provenance offers the           
possibility of reducing storage requirements for object documents. Partial       
reconstruction of documents between specific time points would also reduce the   
computational requirements for the analysis of object documents.                 
                                                                                 
\subsection{A Template for Non-repudiable Evidence}
\label{subsec:nr-evidence}

\begin{figure}[!ht]
\centering
\includegraphics[width=0.85\textwidth]{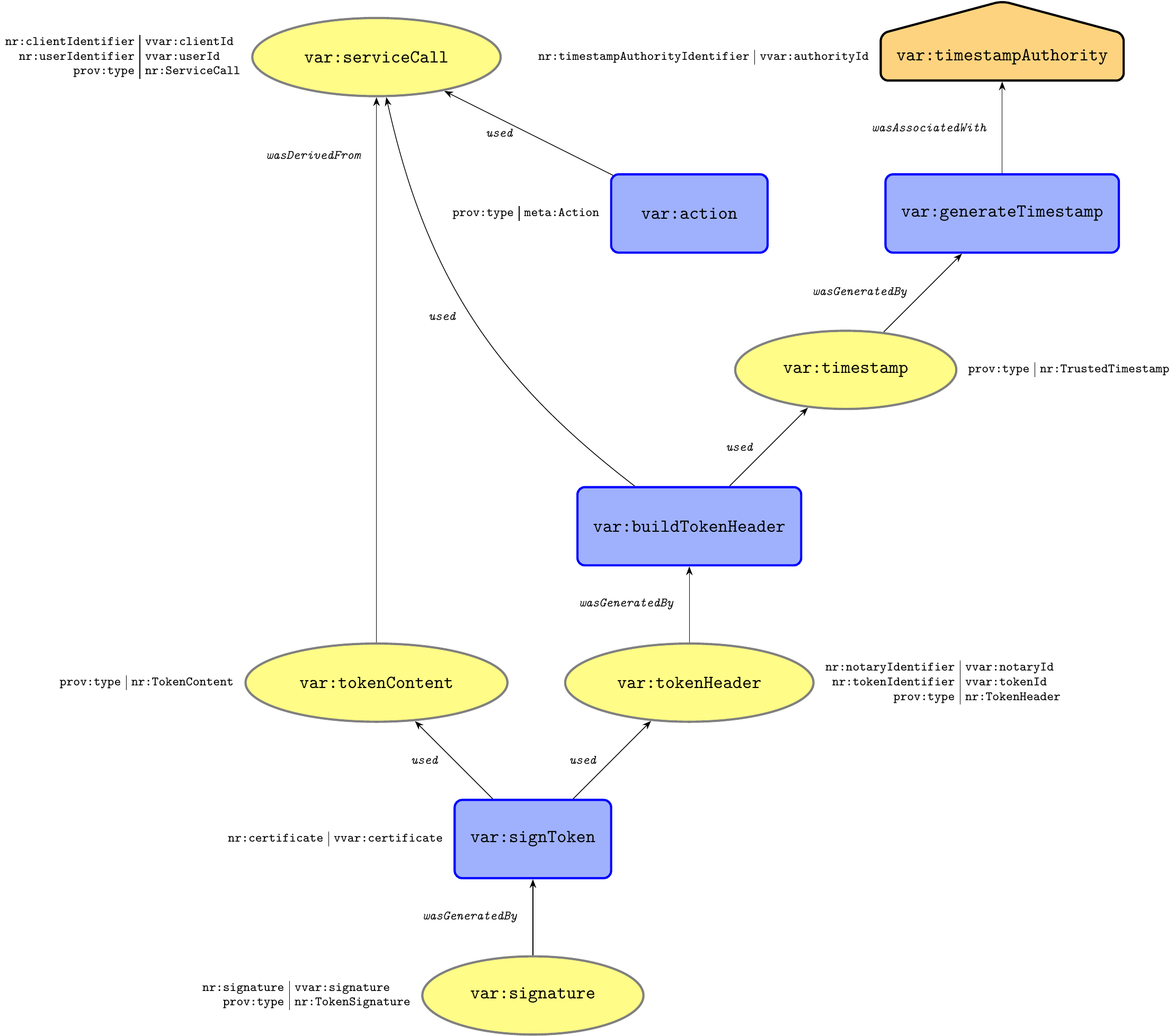}
\caption{Template for capturing evidence required for non-repudiation} 
\label{fig:template-nonrepudiation}
\end{figure}

An important requirement of our solution is the capture of the evidence required to achieve
the non-repudiation of the provenance data being recorded.
The meta-provenance data recorded by each template instantiation within a history document
as described in Subsection~\ref{subsec:metaprovenance} forms the core of this evidence.
However, extra data concerning the security and cryptographic operations later carried
out upon each addition to the meta-provenance record must also be stored in order for
the evidence to provide a guarantee of non-repudiation.
We record this extra data using the template shown in Figure~\ref{fig:template-nonrepudiation}.

The \texttt{var:action} activity represents the document management action
described in Section~\ref{subsec:metaprovenance} and is instantiated with
the same identifier as used for the instantiation of the activity element of the
meta-provenance template corresponding to that action.
It thus serves as the graft point in the history document
between the meta-provenance data recorded about the action and its evidentiary form
and required security attributes, as given by the evidence template.
The \texttt{var:serviceCall} entity represents the service call that requested the
document management action, and holds information about the client application
and user that made the request.
The \texttt{var:tokenHeader} entity represents the meta-data required for
achieving non-repudiation of the service call and the consequent
management action. The token header contains a trusted timestamp, which is
generated by a trusted timestamping authority. The header is identified by a
unique system-generated identifier. 
The \texttt{var:tokenContent} entity holds a representation of the associated
management action, corresponding to the provenance data generated by the meta-provenance module.
The \texttt{var:signature} entity represents the digital signature of the required evidence,
generated by the \texttt{var:signToken} activity. The certificate needed to
verify the signature is given as the value of \texttt{meta:certificate}.

\section{A Non-repudiation Architecture for Decision Support}
\label{sec:nr-architecture}

The goal of the proposed solution is to provide a patient with irrefutable evidence
that a recommendation made by the \gls{decision support system} and the provenance data recording
the generation of that recommendation
originated from the \gls{decision support system}.
There are four actors present in our solution:
\begin{enumerate}
    \item \textit{A patient} using the \gls{decision support system}
    during their treatment
    process and acting on diagnostic recommendations generated by the system.
    \item \textit{A decision support system} consisting of client and server applications,
    and an instance of the provenance template service, provided by and referred to below as the
    \textit{provenance server}.
    A patient will use the client application to request recommendations from the server-side,
    which in turn may request one or more provenance template
    actions to be carried out by the provenance server, for which it will generate
    non-repudiable evidence that is later returned to the patient together with the recommendation.
    \item \textit{A trusted timestamping authority service} which provides signed 
    timestamps. These timestamps are included within the required evidence
    generated to confirm its existence from a specific instant in time.
    \item \textit{A trusted notary service} which stores the required evidence. If a dispute
    about the origin of the provenance data later arises, the notary may be queried to
    either support or contradict a particular claim. 
\end{enumerate}

In the remainder of this section we describe how we use the provenance evidence model given
in Section~\ref{sec:nr-model} to design an extended \gls{decision support system} architecture
that meets the policy requirements outlined in Section~\ref{sec:policy}
and thus ensures non-repudiation of the recommendation-making process.
\subsection{Non-repudiable evidence generation and recording process}

\begin{figure}
\centering
\includegraphics[width=0.85\textwidth]{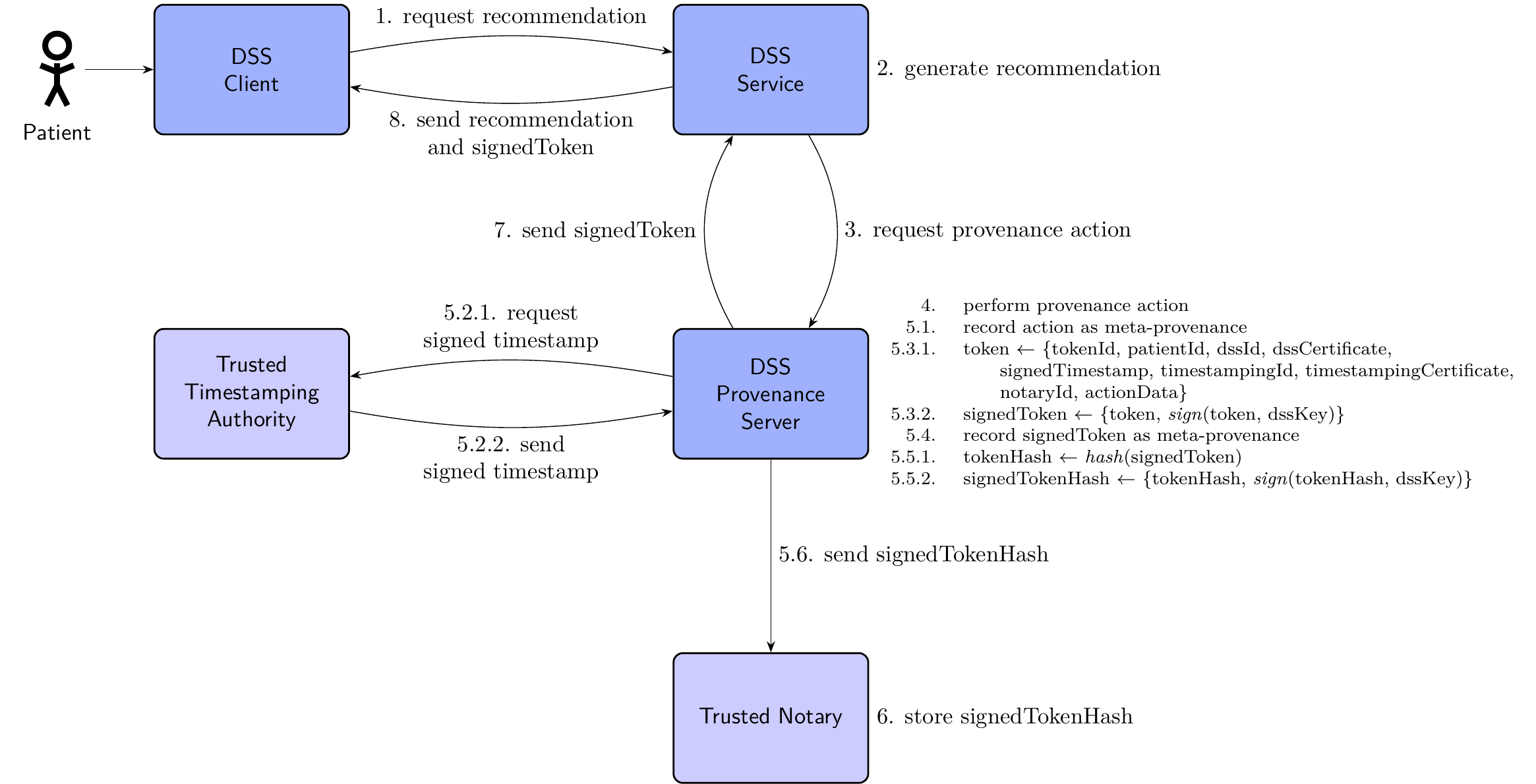}
\caption{The non-repudiable \gls{decision support system} recommendation process} 
\label{fig:evidence-generation}
\end{figure}

The proposed architecture is illustrated in Figure~\ref{fig:evidence-generation}.
The process begins with a request from a \gls{decision support system} client to the
\gls{decision support system}
service (1) that initiates the generation of a recommendation by the system (2).
The recommendation-making process will then in turn request one or more provenance actions to be
performed by the provenance server (3).
The provenance server first executes the requested action upon the object-level document as usual (4).
In the standard architecture this is the point at which the \gls{decision support system} would simply
return the recommendation to the user.
In the extended architecture however
it then records the action as part of the meta-provenance record for the document (5.1)
as described in Subsection~\ref{subsec:metaprovenance}.

The provenance server then constructs a non-repudiation token (5.3.1).
The token is made up of two parts, the \textit{token header} and the \textit{token payload}.
The header contains a system-generated identifier for the token,
an identifier for the patient who made the request,
an identifier for the \gls{decision support system} service,
a \textit{signed timestamp} requested by the server from a trusted timestamping authority (5.2.1),
an identifier for that timestamping authority,
an identifier for the trusted notary that the evidence is to be later sent to for storage,
and the digital certificates required to verify
all signatures within the token.
The payload consists of data representing the meta-provenance recorded
for a provenance action within the recommendation generation process.
In the case of a \textit{registerTemplate} action, this data includes
a hash of a normalised and ordered representation of the template used.
Similarly, in the case of generation actions, the data includes a hash of an
ordered representation of the substitution data used.
The token contains all the information required to resolve potential disputes and addresses policy
requirements \ref{req:token-header}, \ref{req:patient-id} and \ref{req:additional-timestamps}.
The use of a trusted timestamping authority addresses policy requirement~\ref{req:tsa}.

The token is then signed using a private key belonging to the server (5.3.2).
This addresses policy requirement~\ref{req:generation}. 
The signed token is now stored as an instance of
the non-repudiable evidence template shown in Figure~\ref{fig:template-nonrepudiation},
within the appropriate meta-provenance record for the requested
provenance action (5.4).
The signed copy of the token will later be provided to a patient as the irrefutable evidence
of the recommendation generation process.
Since the provenance data used as evidence contains sensitive
information about a patient's health and this information should thus not be provided
directly to the notary, we store a signed hash of the signed token only.
The server therefore now generates a hash of the signed token (5.5.1) and signs that hash
with a second private key (5.5.2).
This second signature is important for preventing the generation of false hashes.
The signed hash is then sent to the trusted notary (5.6) where it is stored (6).

A copy of the signed token for the provenance action is now returned to the
\gls{decision support system} (7). The service then returns the generated recommendation
along with all signed tokens created during recommendation process to the client (8).
The identifier of each signed token is generated by the \gls{decision support system}
and is included in the token header, so that a patient can
access the stored signed token from the meta-provenance record whenever they desire.
This reflects the current movement in healthcare whereby patients are custodians
of their own data~\cite{GORDON2018224}
and addresses policy requirements~\ref{req:backup} and \ref{req:patient-storage}.

\subsection{Non-repudiable Evidence Verification Process}

Once evidence is presented to a patient, according to our policy, its validity must then be checked.
The patient or an authorised agent acting on their behalf, must verify both the integrity and authenticity of the evidence, which consists of the token and its signature.
First, they must verify the content of the token. In particular, they need to ensure that the token payload contains the correct information relating to their medical condition, that the timestamps included in the token are valid, and that there is a link in the token payload to their identity; that is, that they are the correct recipient of the evidence. They must then retrieve the token identifier from the token and look up the corresponding record in the trusted notary. They need to check whether the digital signature of the stored hash was created by the \gls{decision support system}, that it is valid, and that the stored hash is the same for the token that was received from the \gls{decision support system}. Checking the validity of the signatures involves not only verifying the correct computation of the signatures,
but also checking the revocation and expiration information of the keys, the size of the keys and the algorithms used to make sure that the signatures are secure. If these checks succeed, the patient has complete certainty that the provenance data is authentic and that there is a witness (the notary) confirming that fact. If any inconsistency is detected, or one of the digital signatures is invalid, this fact must be reported to the \gls{decision support system} and the patient must not follow the recommendation provided alongside the evidence.

\section{Implementation}
\label{sec:implementation}

\begin{figure}[!ht]
\centering                                                                       
\includegraphics[width=0.85\textwidth]{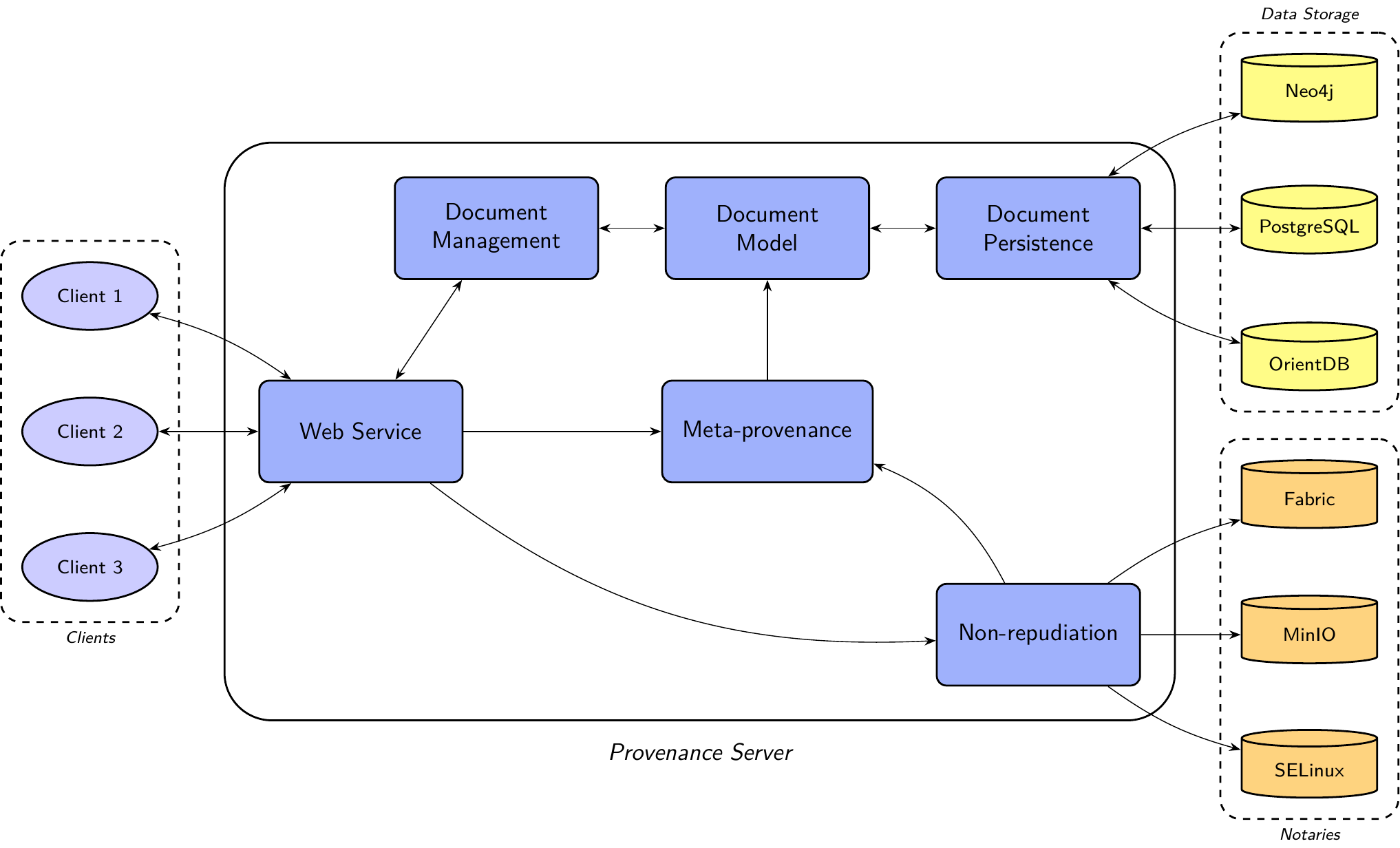}
\caption{The architecture of the provenance server}                              
\label{fig:server-architecture}
\end{figure}

We have implemented a prototype version of the non-repudiation architecture described
in Section~\ref{sec:nr-architecture}.
This involved the development of two new modules for the provenance server.
The first, a meta-provenance module,
carries out the construction of meta-provenance records as detailed in Subsection~\ref{subsec:metaprovenance}.
The second, a non-repudiation module, performs the necessary security and cryptographic operations
upon each addition to a meta-provenance record, generates the necessary non-repudiable evidence,
and appends it to the stored trace.  The functionality of both modules is controlled
by the web service during the handling of relevant incoming requests to the server. 
The use of both modules is optional, however, note that whilst
meta-provenance may be generated without the addition of non-repudiable evidence,
the opposite is not true.
Hashing and signing operations are performed using reference implementations
of algorithms compliant with the PKCS 11 API standard~\cite{pkcs11}, addressing policy requirement~\ref{req:trusted-hw}.
The prototype does not currently make use of a trusted timestamping authority, and instead
generates timestamps locally. This will however be implemented in the near future,
following the RFC 3161 standard~\cite{rfc3161}.
The overall architecture of the server in the context of this paper      
is shown in Figure~\ref{fig:server-architecture}. The remainder of the components
are described briefly below.  

The document model is graph-based and supports both the OPM~\cite{opmSpec2011} and
PROV~\cite{w3cprov2011} provenance specifications.
The document persistence module defines an interface for storing documents, and
provides a number of backends.  The model was designed with graph 
databases in mind, and our core storage backend is Neo4j (\url{https://neo4j.com}),
but we also support those that follow the Tinkerpop (\url{https://tinkerpop.apache.org})
standard, and provide a baseline relational implementation.
The construction of documents is controlled through the document management module,
which is accessed as a web service.

As shown in Figure~\ref{fig:server-architecture}, our prototype system provides three trusted notary implementations, each of which exhibits the required properties of data immutability and data auditability, and meets policy requirement~\ref{req:tamperevident-storage}. Our chosen implementations are a distributed ledger, with hashed blocks and a public ledger (\textit{Hyperledger Fabric}, \url{https://www.hyperledger.org/projects/fabric}); a single store object service, with single-write functionality and object access (\textit{MinIO}, \url{https://min.io}); and a file secured by an append-only access control policy (\textit{SELinux}, \url{http://www.selinuxproject.org}). Our notaries are accessed as a web service, providing calls to add, and validate the presence of, data within each notary\footnote{\url{https://github.com/kclhi/nr}}.

\section{Evaluation}

We hypothesise that each notary implementation is likely to affect the performance of the provenance server differently, depending on the domain within which the server is deployed. Therefore, by providing multiple implementations, our aim is to allow a user to select the notary that works best with the server within a given domain. In order to produce a set of heuristics for notary selection, we now examine the performance of the server when attached to each notary in three use cases with distinct characteristics for the construction of provenance information. These use cases exist within the \consult architecture, a \gls{decision support system} designed to support stroke patients in self-managing their treatments \cite{Chapman2019}.

\begin{figure}
    \centering
    \scalebox{0.67}{\input{diagrams/rules}}
    \caption{Sample of rules used to capture provenance data in the recommendation service}
    \label{fig:rules}
\end{figure}

At the core of the \consult system is an \textit{argumentation}-based recommendation service, which takes facts about a patient and their preferences, and, using a computational form of clinical guidelines, determines a treatment path for them \cite{Kokciyan2018}. This first use case is characterised by a high computation time. In this instance, provenance data provides the aforementioned insight into the recommendations provided by this service. To extract this data, we augment the service's rule-base with an additional set of rules, which are satisfied when the system makes particular decisions, and thus output the required provenance data. We structure these rules in the style of \cite{Toniolo2014}, and an example is given in Figure~\ref{fig:rules}. 

The facts used by the recommendation service are based upon sensor data (and a patient's EHR), which the \consult system gathers from wearable devices. This second use case is characterised by high volumes of data. Here, provenance data is useful for auditing purposes, for example to aggregate sensor data and identify erroneous readings, before subsequently tracking them back to the device from which they originate. To extract this data, we examine the sensor readings that arrives at a central service in the \consult system. 

To interface with the \consult system, users engage with a chatbot. This chatbot is able to provide the patient with healthcare information, which includes the information provided by the recommendation service, although we do not consider this interaction as a part of this third use case. This aspect of the system is characterised by its non-determinism, as we cannot know, prior to the execution of the chatbot, which answers a user will provide. Much like the recommendation service, in this situation provenance data provides insight into the decisions made by the chatbot, and various parts of the \consult chatbot logic are augmented to extract this data. 

The output from each of these use cases is used as the basis for constructing substitutions for a set of templates designed for the \consult \gls{decision support system}. These templates capture the key agents (e.g. a patient), entities (e.g. a sensor reading or a clinical guideline) and activities (e.g. the generation of a recommendation), in the \gls{decision support system}. In the case of the chatbot, each substitution is constructed and submitted incrementally as zones, as the interaction with the chatbot progresses.

\subsection{Experiments and Results}

We now examine the performance of the provenance server when attached to each notary, within each of these use cases. To do so, we further augment the \consult system in order to simulate patients interacting with each use case. The results of these simulations are shown in Table \ref{table:results}, which reflects the average response time, and related statistical tests, of each call to the server from \consult, over 1000 simulations ($N = 1000$).
Note that these experiments were performed before the completion of the prototype
but still offer a relative comparison of performance.

\input{tables/results.tex}

Examining first the performance of different notaries against one another, we note that the introduction of a ledger-based notary results in the most significant overhead in terms of response time, which is to be expected, given that speed is a common criticism of the technology. While this may make a ledger appear to be the least attractive option, it may still be a consideration when deploying the provenance server, as the use of a ledger brings additional benefits, such as decentralisation, which may outweigh the impact of an increased response time. Of the remaining two notaries, the use of single-write object store offers the best performance, over an append-only file, which is interesting given the low-level nature of the latter, and suggests that, when linked to our server, a notary technology optimised for the storage of client data is more efficient.

In terms of the performance of the same notary across different deployment domains, we note that, in addition to offering the best performance, the use of a single-write object store also guarantees consistent performance when operating in domains with differing provenance data collection properties. This may make the object store an attractive option in domains with uncertain properties. In contrast, while responding broadly consistently to the high throughput of data found in the sensor use case, and when working with the complexity of the rule-based recommendation service, both the ledger and file offer improved performance in the chatbot use case. While, in general, this is to be expected, given that a number of smaller submissions are made to the server during the incremental construction of resources, rather than a single larger submission, it is interesting to note that this style of resource construction has the most significant impact on the performance of these two notaries.

\section{Conclusions and Future Work}

We have identified the importance of the role of provenance data
within a \gls{decision support system}, as evidence designed
to be used in the resolution of potential disputes about the actions of the system.
We have shown that a \gls{decision support system} -- the evidence generator -- has a motive to disrupt this evidence in order to protect its interests, and thus
the authenticity and integrity of the evidence must be established. That is, provenance data used as evidence
must exhibit non-repudiation of origin.
In order to achieve this goal, we have defined a security policy for non-repudiable
evidence in the context of a \gls{decision support system},
developed a fully general provenance-based model to represent such evidence,
and then proposed an extended \gls{decision support system} architecture that 
meets the requirements of our policy.

Our solution allows us to present to the user a comprehensive survey of all provenance actions taken
by the system on their behalf and to retrieve and validate the authenticity of that data at any point thereafter.

We have developed a prototype implementation of our architecture, by extending
our provenance template service with the functionality described in our evidence model, and
creating a web service interface for trusted notary applications.
We evaluated the performance of the prototype using three contrasting provenance generation
use cases arising within the \consult \gls{decision support system}.

Following further improvements to our prototype, we now intend to investigate how our
model may be extended to work within a distributed environment, in which provenance data is being
generated by multiple systems working in multiple domains.

\bibliographystyle{splncs04}
\bibliography{references}

\end{document}

%% file: glossary.tex
\newglossaryentry{decision support system}{%
  name={DSS},
  description={NA}
}

%% file: diagrams/rules.tex
\begin{lstlisting}
 a(provenance(source(giveRecommendation), relationship(wasAssociatedWith), target(Patient))) :-
  a(aspt([goal(G), action(A), promotes(A,G)], action(A))), patient(Patient).
 a(provenance(source(giveRecommendation), relationship(used), target(S))) :-
  suffers_from(Patient,S).
\end{lstlisting}

%% file: tables/results.tex
\begin{table}[!t]
    \centering
    \begin{tabular}{l|l|l|l}
       \textit{} & Recommendation & Sensor & Chatbot \\
       \hline
       Ledger & 7.04 (.05,.05; .05,.05) & 6.25 (.05,.05; .05,.05) & 3.02 (.05,.05; .05,.05) \\
       Object & 0.56 (.05,.05; .6,.4) & 0.56 (.05,.05; .6,.6) & 0.55 (.05,.05; .4,.6) \\
       File & 1.17 (.05,.05; .05,.05) & 1.36 (.05,.05; .05,.05) & 0.82 (.05,.05; .05,.05) \\
    \end{tabular}
    \caption{Average response time of provenance server (seconds) using different notaries in different \consult use cases, and associated (maximum) \textit{p}-values (other notaries; other use cases).}
    \label{table:results}
\end{table}